\documentclass[iop]{emulateapj}

\usepackage{natbib}
\bibpunct{(}{)}{;}{a}{}{,}                        

\shorttitle{The Next Generation Transit Survey Prototyping Phase}
\shortauthors{McCormac et al.}

\begin{document}

\title{The Next Generation Transit Survey - Prototyping Phase}

\author{J. McCormac\altaffilmark{1, 2}, D. Pollacco\altaffilmark{1, 2}, P. J. Wheatley\altaffilmark{1}, R. G. West\altaffilmark{1, 3}, S. Walker\altaffilmark{1}, J. Bento\altaffilmark{6, 1}, I. Skillen\altaffilmark{5}, F. Faedi\altaffilmark{1, 2}, M. R. Burleigh\altaffilmark{3}, S. L. Casewell\altaffilmark{3}, B. Chazelas\altaffilmark{4}, L. Genolet\altaffilmark{4}, N. P. Gibson\altaffilmark{2}, M. R. Goad\altaffilmark{3}, K. A. Lawrie\altaffilmark{3}, R. Ryans\altaffilmark{2}, I. Todd\altaffilmark{2}, S. Udry\altaffilmark{4}, C. A. Watson\altaffilmark{2}}
\altaffiltext{1}{Department of Physics, University of Warwick, Coventry CV4 7AL, UK}
\altaffiltext{2}{Astrophysics Research Centre, School of Mathematics and  Physics, Queen's University, Belfast, BT7 1NN, UK}
\altaffiltext{3}{Department of Physics \& Astronomy, College of Science \& Engineering, University of Leicester, University Road, Leicester, LE1 7RH}
\altaffiltext{4}{Observatoire Astronomique de l`Universit{\'e} de Gen{\'e}ve, 51 ch. des Maillettes, 1290 Sauverny, Switzerland}
\altaffiltext{5}{Isaac Newton Group of Telescopes, Apartado de Correos 321, E-38700, Santa Cruz de la Palma, Spain}
\altaffiltext{6}{Research School of Astronomy and Astrophysics, Mount Stromlo Observatory, Australian National University, Cotter Road, Weston Creek, ACT, 2611, Australia}
\email{j.j.mccormac@warwick.ac.uk}
\submitted{Accepted PASP 2016 Oct 28}

\begin{abstract}
We present the prototype telescope for the Next Generation Transit Survey, which was built in the UK in 2008/09 and tested on La Palma in the Canary Islands in 2010. The goals for the prototype system were severalfold: to determine the level of systematic noise in an NGTS-like system; demonstrate that we can perform photometry at the (sub) millimagnitude level on transit timescales across a wide field; show that it is possible to detect transiting super-Earth and Neptune-sized exoplanets and prove the technical feasibility of the proposed planet survey. We tested the system for around 100 nights and met each of the goals above. Several key areas for improvement were highlighted during the prototyping phase. They have been subsequently addressed in the final NGTS facility which was recently commissioned at ESO Cerro Paranal, Chile. 
\end{abstract}


\section{INTRODUCTION}

\label{sec:introduction}

The Next Generation Transit Survey (NGTS\footnotemark\footnotetext{www.ngtransits.org}) is a new wide-field photometric survey for transiting exoplanets recently commissioned at ESO Paranal, Chile. The motivation behind NGTS is to understand planetary formation and evolution, as well as determining atmospheric and bulk compositions of Neptune and super-Earth sized exoplanets. This can only be achieved by characterising, both photometrically and spectroscopically, a large sample of such objects. The \emph{Kepler} survey has shown there to be an abundance of Neptune sized objects in their field \citep{2013ApJ...766...81F}. However, while the results from \emph{Kepler} are important in terms of statistics, the mean brightness of host stars of small planets observed by \emph{Kepler} is too faint for efficient spectroscopic follow up from the ground. The NGTS project (Wheatley et al., in prep, \citealt{2013EPJWC..4713002W}, \citealt{2012SPIE.8444E..0EC}) was designed to detect this population of Neptune-sized transiting exoplanets around bright (V$\leq$13) K and early M-type stars accessible to the ESO Telescopes.

\citet{2006MNRAS.373..231P} showed that systematic (red) noise is non-negligible when searching for transiting exoplanets and hence limits the detection efficiency of most ground-based surveys. Initial predictions of transiting planet detection rates considered uncorrelated (white) noise only, overlooking the effects of red noise. Pont et al. suggest that this may explain the inconsistency between transiting exoplanet detection rates compared to initial predictions. \citet{2006MNRAS.373.1151S} investigated the effects of systematic noise on the detection efficiency of SuperWASP \citep{2006PASP..118.1407P} and concluded that the survey suffered significantly from a systematic noise component in the data. The noise in SuperWASP was determined to be a combination of both red and white (pink) noise. The level of systematic noise was reduced using the SysRem detrending algorithm \citep{2005MNRAS.356.1466T}, but was not removed completely. As a result SuperWASP was forced to acquire more transits of each planet to increase the signal-to-noise ratio of their detections. 

In order to meet the scientific goals of the NGTS project it is crucial we understand the type of noise in an NGTS-like system and identify any possible sources of systematic noise. A prototype telescope for NGTS (hereafter NGTS-P) was constructed at Queen's University Belfast in the winter of 2008 and spring of 2009. NGTS-P was tested at the Observatorio del Roque de los Muchachos (ORM) on La Palma during 2009 and 2010, conducting several experiments to determine the technical feasibility of the proposed NGTS project. The main aim of NGTS-P was to determine the level of systematic noise in an NGTS-like system and prove that millimagnitude photometry or better could be achieved on transit time scales using wide-field Newtonian telescopes of modest apertures, coupled with scientific grade CCD cameras with deep depletion silicon. The use of deep depleted (DD-) CCDs, with their increased quantum efficiency at red wavelengths, was proposed to increase the sensitivity of the survey to K and M-type stars. This in theory allows for the detection of smaller transiting exoplanets. Also, using DD-CCDs significantly reduces the effects of fringing, which plague traditional CCD imaging at redder wavelengths. 

This paper is organised as follows: in \S \ref{sec:hardware} we describe the NGTS prototype telescope and control software. In \S \ref{sec:observations} we describe the observations and data reduction process. In \S \ref{sec:performance} we outline the photometric performance of the prototype and give the results from several on-sky tests. In \S \ref{sec:discussion} we discuss the results from the prototyping phase and highlight the areas where improvement was required for NGTS. Finally, in \S \ref{sec:conclusion} we close with our conclusions from the NGTS prototyping phase.

\begin{figure}
\includegraphics[scale=0.27,angle=0]{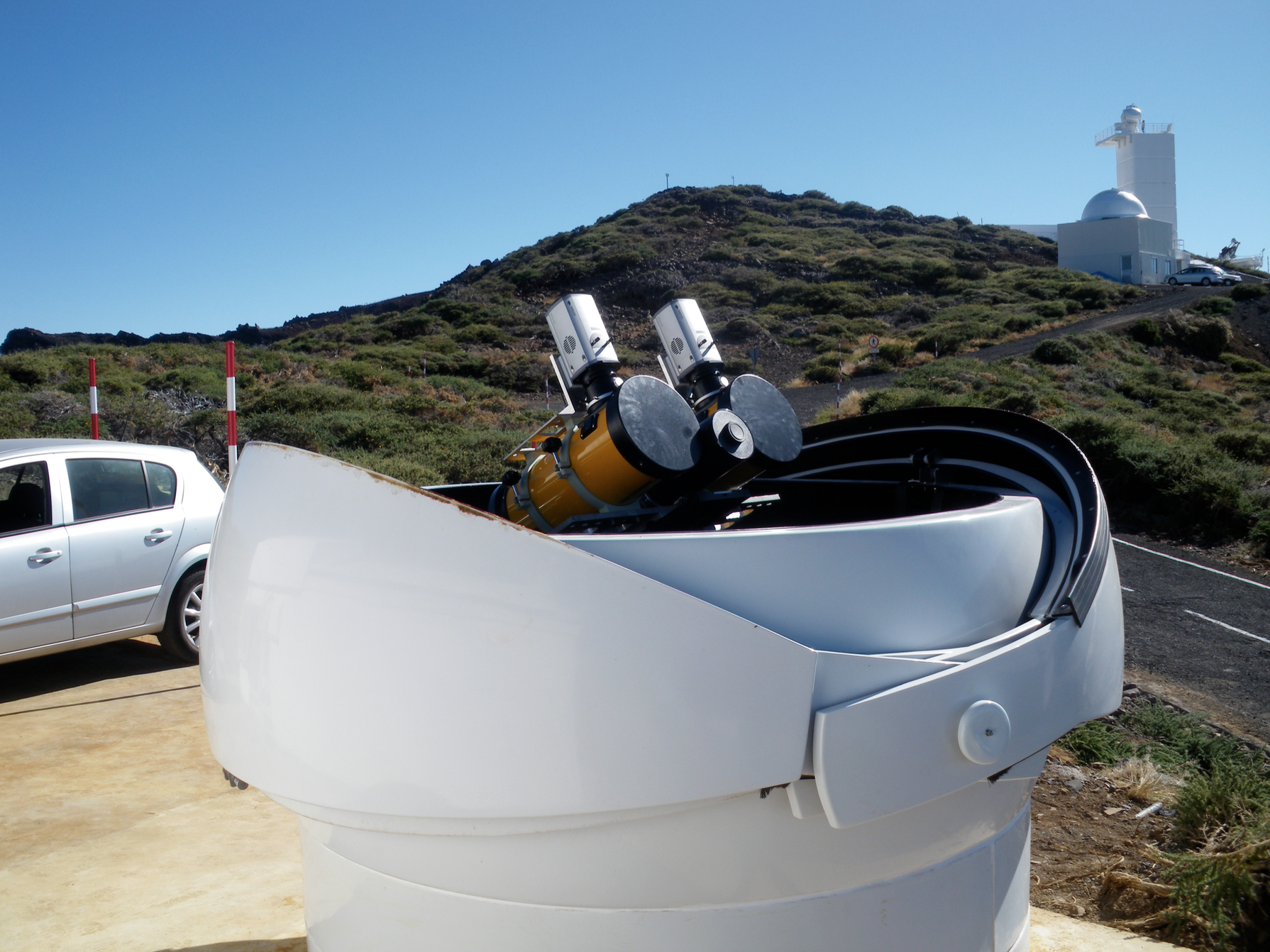}
\caption{NGTS-P at the Observatorio del Roque de los Muchachos, La Palma, Spain. The prototype initially contained two telescopes but the second camera developed a fault and was returned to the manufacturer shortly after commissioning. All tests described above were carried out with a single camera (iKon1).}
\label{figure:prototype}
\end{figure}

\section{THE PROTOTYPE TELESCOPE}
\label{sec:hardware}

NGTS-P consisted of an Andor Technologies\footnotemark\footnotetext{http://www.andor.com} iKon-M 934N BR DD-CCD camera (hereafter iKon1) coupled to a Takahashi\footnotemark\footnotetext{http://www.takahashi-europe.com} E-180 wide-field reflecting telescope. The CCD was back illuminated with $1024\times1024$ pixels made by e2v and had a peak quantum efficiency $>90$\% ($\approx800$ nm). The sensor was made from high-resistance, DD silicon which produces significantly less fringing as more near-infrared photons are captured before interfering with incoming light. The camera was cooled to $-80^{\circ}$C using a five-stage Peltier thermoelectric cooler resulting in a dark current of $2.46$ e$^{-}$ pixel$^{-1}$ min$^{-1}$. A readout frequency of $1$ MHz ($10^{6}$ pixels s$^{-1}$) was chosen as a compromise between speed and noise. This camera setting gave a gain, read noise and readout time of $1.2$ e$^{-}$ ADU$^{-1}$, $7.1$ e$^{-}$ and $1.05$ seconds, respectively. See Fig \ref{figure:prototype} for a photo of NGTS-P installed at the ORM on La Palma. 

\begin{figure}
\includegraphics[width=3.5in, trim=15mm 0mm 5mm 18mm, clip]{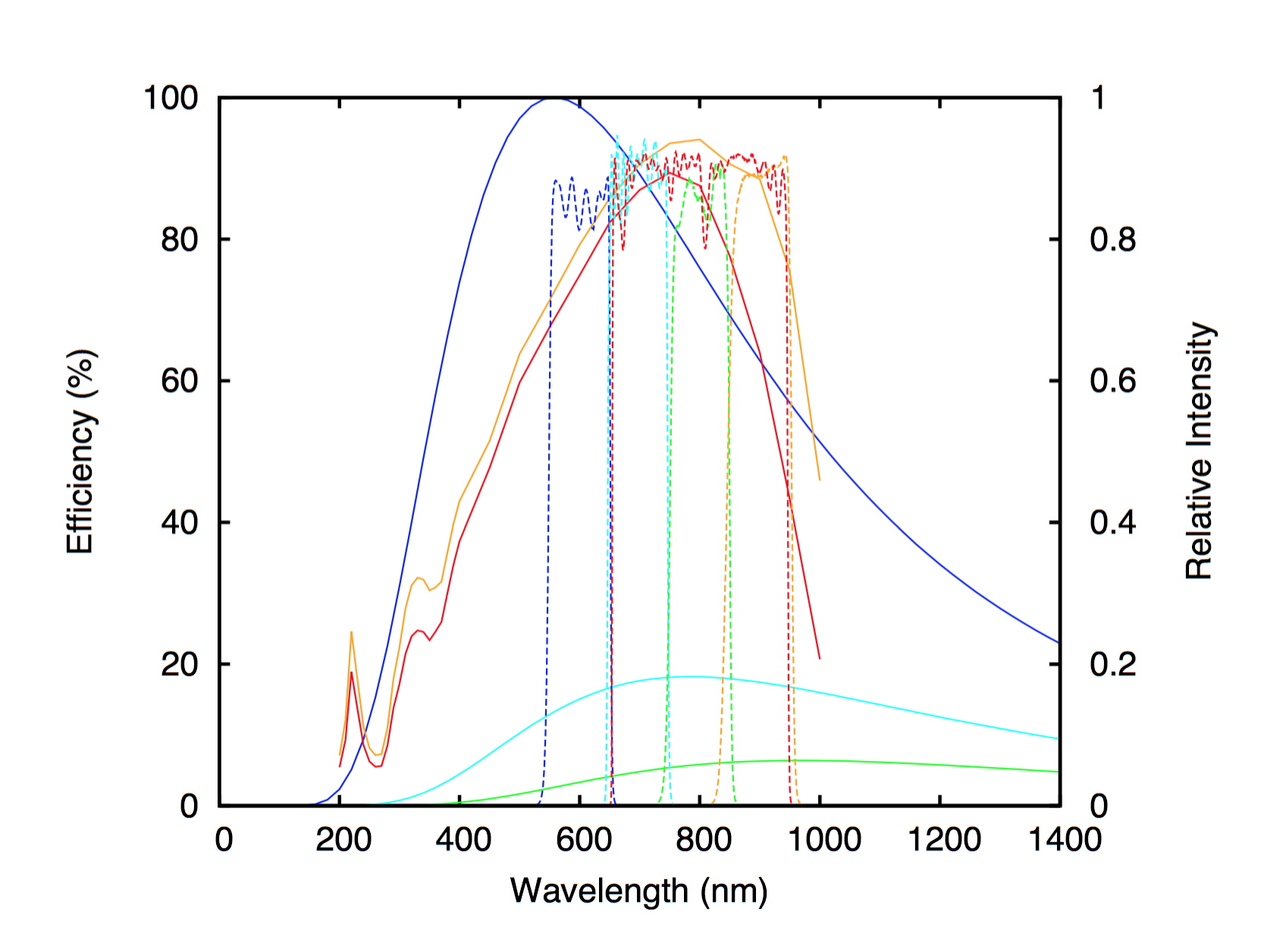}
\caption{Andor iKon-M 934N BR-DD quantum efficiency curve (T$_{\mathrm{CCD}}=25^{\circ}$C solid orange, T$_{\mathrm{CCD}}=-90^{\circ}$C solid red) over plotted with the blackbody emission curves of a K0, K9 and M4.5 star and the NGTS custom bandpass filter set (dotted blue, cyan, green, orange and red lines). Peak emission wavelengths ($558$, $784$ \& $967$ nm for K0 (solid blue), K9 (solid cyan) \& M4.5 (solid green) stars, respectively) for these stellar classes are well matched to the sensitivity of the DD-CCDs. The intensities given by Planck's energy distribution equation have been scaled to $1$ for clarity. }
\label{figure:filtersqebbcurves}
\end{figure}

The Takahasi E-180 is an optically fast ($f/2.8$) Newtonian telescope with an aperture and focal length of $180$ and $500$ mm, giving a field-of-view and plate scale of $1\fdg56\times1\fdg56$ and \mbox{$5\farcs36$ pixel$^{-1}$,} respectively. A single fixed filter was housed in the camera-to-telescope coupling. A set of $5$ bandpass filters ($4$ narrow, $1$ broadband) were available for NGTS-P. In order to characterise the camera's broadband red performance all of the tests described in \S \ref{sec:performance} were carried out using the pseudo-$I$ band filter ($650-950$ nm, see Fig \ref{figure:filtersqebbcurves}, red dashed line). Using the pseudo-$I$ band filter allows us to better match the spectral energy distribution of K and early M-type stars, without being unnecessarily restrictive to bluer photons. 

The telescope was mounted on a Paramount ME\footnotemark\footnotetext{http://www.bisque.com} German Equatorial Mount (GEM) and auto-guided using a separate refracting telescope and auto-guider camera. One drawback of using a GEM is the requirement to flip the mount when crossing the meridian. This reverses the position of all the stars on the CCD causing additional complications when conducting precise photometry. Therefore, to avoid this all of the tests described in \S \ref{sec:performance} were conducted on one side of the meridian only. We note that a GEM was used in the prototyping phase in order to be cost effective; such a mount was never envisaged for the full NGTS facility. 

NGTS-P was housed inside a $7$ ft Astrohaven\footnotemark\footnotetext{http://www.astrohaven.com} clamshell dome and the focus of each telescope was controlled using a Robofocus\footnotemark\footnotetext{http://www.robofocus.com} absolute stepper motor. The observatory was controlled using high level scripts written in the C programming language. These scripts controlled the camera using the Andor Software Development Kit and the telescope mount via TheSky6 interface from Software Bisque. An additional script controlled the dome and monitored the weather conditions at SuperWASP-N, beside which NGTS-P was installed, allowing for a safe shutdown of the telescope during adverse weather conditions. 

\section{OBSERVATIONS \& DATA REDUCTION}
\label{sec:observations}

Observations with NGTS-P were carried out remotely from sea level on La Palma between September 2009 and July 2010, collecting over $100$ nights of data. Each night's data were reduced using the standard routines in IRAF\footnotemark\footnotetext{IRAF is distributed by the National Optical Astronomy Observatories, which are operated by the Association of Universities for Research in Astronomy, Inc., under cooperative agreement with the National Science Foundation.}. A master bias and dark frame was constructed by median combining a minimum of $11$ frames. After some initial analysis it became evident that the flat fielding process was adding non-linear noise to the data, due to scattered light in the flat fields, which could not be removed via detrending (see \S\ref{subsec:optical}). As a result the data analysed here were reduced without flat fielding. The airmass at mid-exposure was calculated using the \emph{setairmass} routine in IRAF and the Julian Date (JD) at the mid point of each observation was converted to both the Heliocentric Julian Date (HJD) and the Barycentric Julian Date (BJD\_TDB) using the method of \citet{2010PASP..122..935E}.  Fields for NGTS-P were chosen to allow for $\sim4$ hours of observations without requiring a pier flip. Several test fields were observed so long as they avoided very crowded regions near the galactic plane, the moon and extremely bright stars. To characterise the noise properties of NGTS-P required observing an average sample of stars, therefore the telescope was able to point semi-arbitrarily. An exposure time of $30$ s was typically used. With hind-sight this was too long. The exposure time for the final NGTS facility has been decreased to $10$ s to permit photometry of brighter stars and account for the increased efficiency of the final system as well as the $19$\% increase in telescope collecting area compared to NGTS-P (20cm vs 18cm diameter).

Stars were identified in the first frame of each run using SExtractor \citep{1996A&AS..117..393B}. Those with peak flux $>45\,000$ e$^{-}$ and those within $50$ pixels of the edge of the CCD were ignored. Light curves were extracted from the series of images using the DAOPHOT package \citep{1987PASP...99..191S} in IRAF.  In the first frame aperture photometry was performed on positions found by SExtractor and in subsequent frames the internal centring algorithm of DAOPHOT was used to track the drift of the stars over time. The raw light curves for each night were analysed and those of variable stars or with errors from the photometry, such as skipping of the aperture to brighter adjacent sources during the centroiding process, were excluded. The remaining stars were then passed through our implementation of the SysRem detrending algorithm \citep{2005MNRAS.356.1466T}. Below is a brief outline of the process.

The SysRem detrending algorithm can remove any effect that appears linearly in a large portion of the observed light curves. It acts to minimise the global expression:

\vspace{-0.4in}
\begin{center}
\begin{eqnarray}
S^{2}=\sum_{ij}\frac{(r_{ij}-c_{i}a_{j})^{2}}{\sigma^{2}_{ij}}\;, \label{eq:Tamuz} 
\end{eqnarray}
\end{center}

\noindent by iteratively finding the two best sets of $\lbrace c_{i} \: ; i=1,N \rbrace$ and $\lbrace a_{j} \: ; j=1,M \rbrace$ of $N$ light curves from $M$ images, where $r_{ij}$ is the residual flux from the average subtracted light curve $i$ in image $j$. The coefficients $a_{j}$ and $c_{i}$ are per-image ($j$) and per-object ($i$) weights and $\sigma_{ij}$ is the uncertainty on $r_{ij}$. A full description of the detrending algorithm and its application to an OGLE data set is given by \citet{2005MNRAS.356.1466T} and \citet{2007ASPC..366..119M}, respectively. 

A stopping criterion must be defined for SysRem so that only the strongest trends are removed and not the signal from planetary transits etc. The stopping criterion for SysRem was set to remove effects that were stronger than $95$\% of all random effects ($\alpha = 0.95$). The detrending algorithm typically removed two trends, an airmass trend and a second, much less dominant trend, most likely due to colour-dependent atmospheric extinction. In the majority of cases the stopping criterion was almost satisfied after the removal of the airmass trend only, $\alpha \sim 90\%$. This meant that trends stronger than $\sim90\%$ of all random effects had been corrected with the removal of a single trend, highlighting the absence of systematic noise in the system. An analysis of the prototype's systematic noise is given in \S\ref{subsec:systematics}.

\section{NGTS-P PERFORMANCE}
\label{sec:performance}

\begin{figure}
\includegraphics[width=3.5in, trim=15mm 0mm 5mm 5mm, clip]{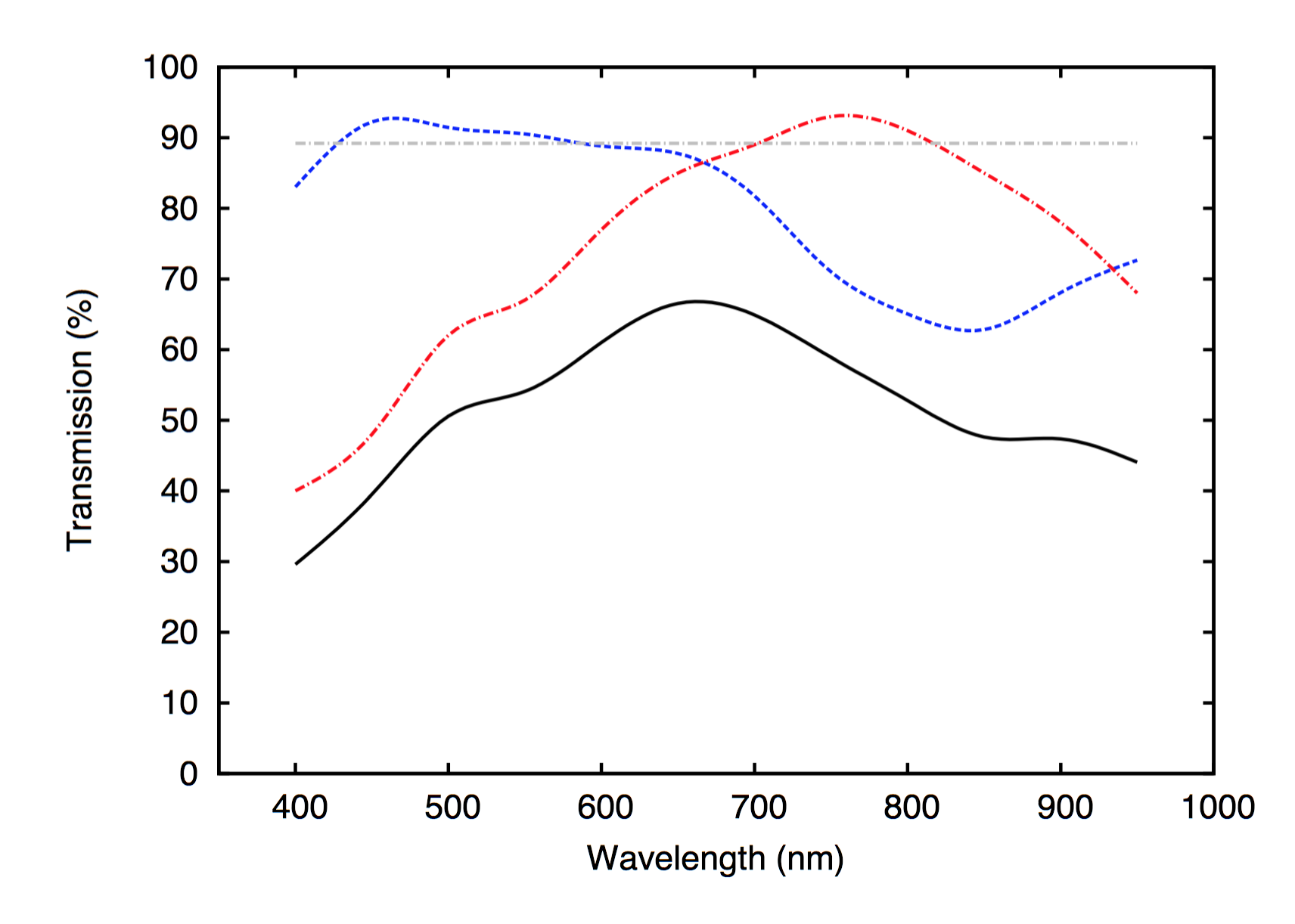}
\caption{NGTS-P optical model. The dashed blue line shows the total semi-theoretical transmission of the Takahashi E-180. The red and grey dot-dashed lines show the QE of iKon1 and the throughput of the NGTS-P broadband pseudo-${I}$ filter. The total optical model is given by the solid black line. However, we note that the throughput model for the Takahashi E-180 may be considerably different beyond $700$ nm.}
\label{figure:OpticalModel}
\end{figure}

\subsection{Tracking and Autoguiding}
\label{subsec:tracking}
We observed several target fields with NGTS-P during September 2009 to investigate the tracking and autoguiding performance of the telescope. We monitored a drift of $3.79$ pixels h$^{-1}$ in the stellar positions on the CCD while autoguiding. After extensive investigation the source of the flexure remained elusive. It most likely stemmed from the significant ($\sim30$ cm) mechanically unsupported back focal distance required to achieve focus with the separate autoguiding telescope and camera. All of the data presented here therefore contained a slow drift in the stellar positions on the CCD. Our solution to the drifting problem is discussed in more detail in \S\ref{sec:discussion}.

\subsection{Optical}
\label{subsec:optical}

There were a total of $6$ optical elements in the beam of the telescope, the DD-CCD, filter and $4$ elements inside the Takahashi E-180 ($2$ mirrors and $2$ lens elements in the corrector). Limited information ($400-700$ nm) on the spectral response of the mirrors and corrector elements was obtained from Takahashi Europe. Beyond $700$ nm only approximations of the spectral response of the telescope can be made. The reflectance of each mirror beyond $700$ nm was approximated using the spectral response of aluminium.  The relative reflectance of mirror aluminium was measured in the lab and the reflectance curve was calibrated using the data from Takahashi. The maximum reflectance of the Takahashi mirror occurs between $450$ and $500$ nm where $\sim96.25$\% of the light is reflected. The relative reflection curve of mirror aluminium was scaled to match the Takahashi reflectance in this wavelength range. This allowed us to estimate of the mirror response beyond $700$ nm. However, we note that the peak reflectance of uncoated aluminium is typically $<96.25$\%, hinting that the mirrors in the Takahashi E-180 might be coated and therefore respond differently beyond $700$ nm. It is possible that the real efficiency is less than quoted here.

It was more difficult to estimate the transmission of the corrector beyond $700$ nm without knowing more about the type of glass used and any coatings which may have been applied. In this case a simple extrapolation of the Takahashi data was made from $700$ to $950$ nm. The broadband NGTS-P pseudo-$I$ filter was used for the majority of the observations hence for simplicity its average throughput of $89.2$\% has been applied at all wavelengths in the optical model. iKon1 was most sensitive at $\sim800$ nm with a steep decline in QE beyond $950$ nm. All the optical elements have been combined in 
Fig \ref{figure:OpticalModel} to give a theoretical model of the optical system. The model presented here excludes the $\sim20$\% reduction in the collecting area of the telescope caused by the secondary mirror obstruction. 

\begin{figure}
\includegraphics[scale=0.28,angle=0, trim = 30mm 15mm 0mm 20mm, clip ]{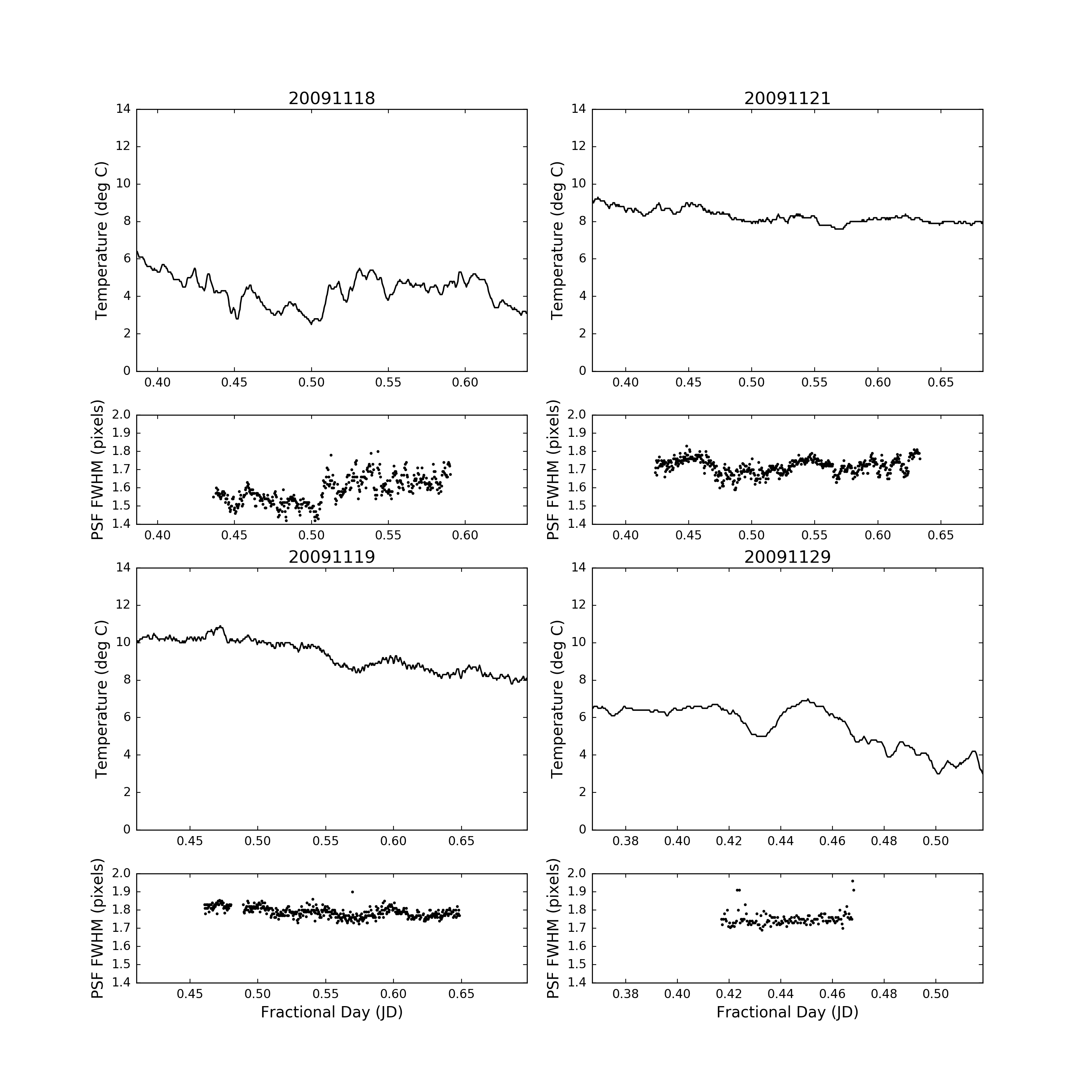}
\caption{Examples of PSF stability of NGTS-P with time and temperature from 4 nights in November 2009. The upper panel in each quadrant shows the ambient temperature and the lower panel shows the variation in the PSF during the observations.}
\label{figure:PSFwTIMEwTEMP}
\end{figure}

To characterise the optical performance of the telescope and camera we monitored the change in PSF with the ambient temperature. The PSF remained suitably constant with time and temperature (see Fig \ref{figure:PSFwTIMEwTEMP}),  however a small correlation between telescope focus and temperature was observed. Although the previous correlation with temperature was within our desired limits, the final telescope design of NGTS uses a carbon fibre structure which further minimises the effects of temperature changes. We also monitored the scattered light content in the flat fields and discovered a significant amount of scattered light in the twilight flats. This was traced to scattered light reflecting from the inside of the telescope and striking the CCD from outside the beam. We installed a lightweight baffle which reduced the effects of the scattered light but finally, as explained in \S\ref{sec:observations}, we decided not to flat field the prototype data due to the additional noise added during the flat fielding process.

As the estimated throughput of the Takahashi E-180 was unsuitably low ($<70$\% between $800-900$ nm) and confirmation of this was unavailable from the manufacturer, a different red-optimised telescope made by Astro System Austria has been implemented in the final NGTS facility. The internal surfaces of the new telescopes are coated with a black flocking material and a $500$ mm long baffling collar has been fitted to each telescope aperture to reduce the impact of scattered light. The solutions to the optical issues with NGTS-P are further described in \S\ref{sec:discussion}.

\subsection{Photometric}
\label{subsec:photometric}
\begin{figure*}
\includegraphics[trim = 0mm 0mm 0mm 0mm, clip, width=7in]{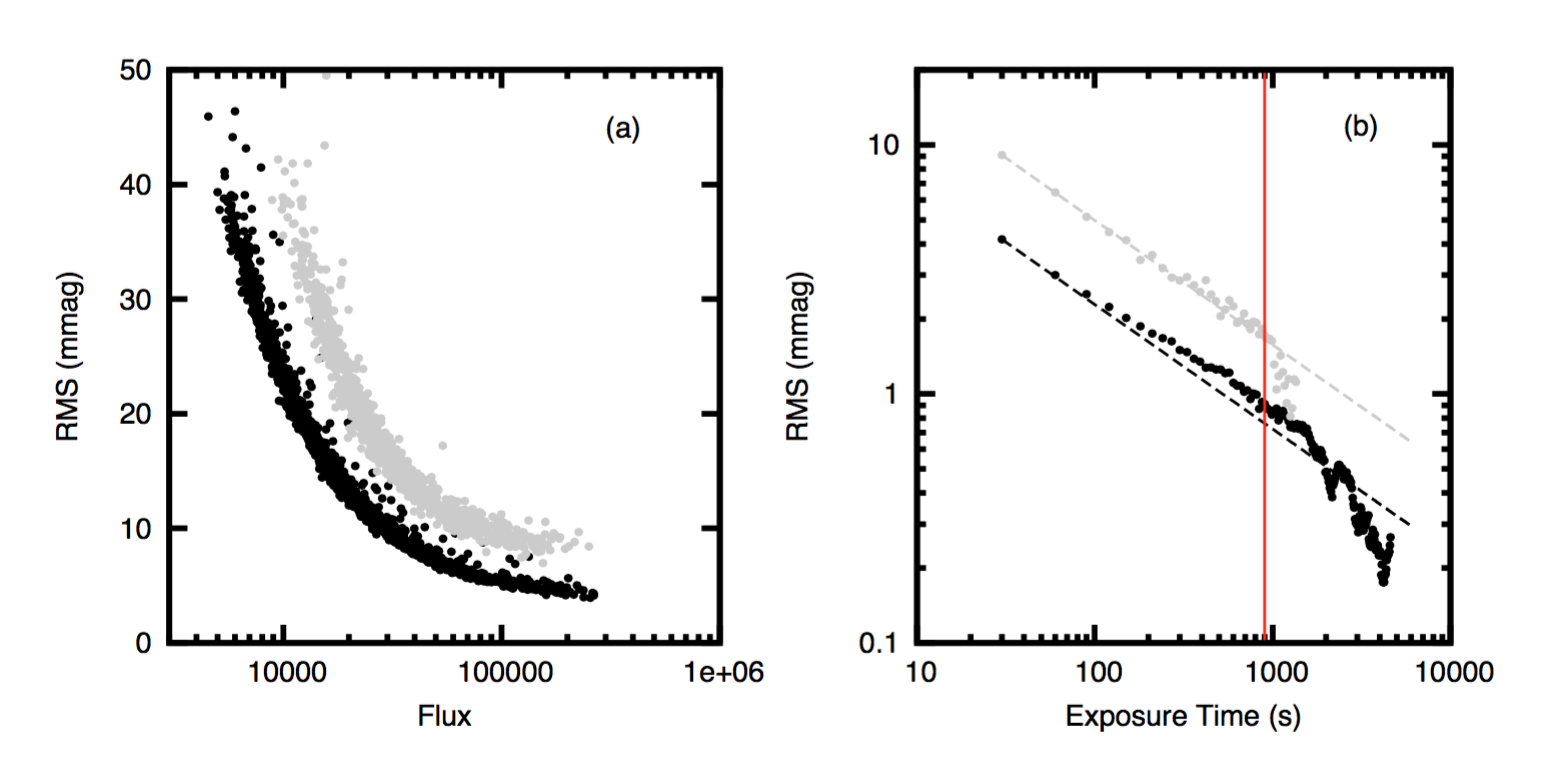}
\caption{Photometric accuracy of NGTS-P. Panel $a$ compares the RMS vs flux plots of a typical full (grey points) and new (black points) moon night. Panel $b$ shows the RMS with binned exposure time for the 9 brightest stars on each night. The dashed lines show the expected $1/\sqrt{n}$ decrease in RMS in the presence of Gaussian noise only, where $n$ is the number of points per bin. The solid red line highlights the $900$ s binned exposure time required to reach $1$ mmag RMS at new moon. The approximate V magnitude range covered here is $10 < V < 14$. The bright limit was defined by the $\sim45,000$ peak count limit on the linear response of iKon1. The faint limit was set conservatively to give a flux level of $\sim5000$ e$^{-}$.}
\label{figure:PhotometricAccruacy}
\end{figure*}

We characterised the photometric performance of NGTS-P by observing a given field from the eastern horizon to the meridian over many nights, with sky brightnesses ranging from full to new moon conditions. The data were reduced using the method outlined in \S\ref{sec:observations} and we created diagrams of the Root Mean Squared (RMS) of the light curve noise vs magnitude and RMS of the light curve noise vs binned exposure time diagrams for each night. The RMS vs magnitude diagram was compared to a theoretical noise model of the telescope and the RMS vs binned exposure time diagram was compared to the $1/\sqrt{n}$ decrease in RMS expected in the presence of Gaussian (white) noise only, where $n$ is the number of points per bin. The results from two typical nights ($1$ during full and $1$ during new moon conditions) can be seen in Fig \ref{figure:PhotometricAccruacy}.  Fig \ref{figure:PhotometricAccruacy} shows that our goal of achieving $1$ mmag photometry of bright sources is possible in $\sim900$ s during new moon conditions, and  $<2$ mmag RMS is achievable on the same time scale during full moon. This was promising given that the tracking and scattered light issues previously highlighted still required addressing.

\subsection{Systematic Noise}
\label{subsec:systematics}
\begin{figure*}
\includegraphics[scale=0.65,angle=0,width=7in, height=4in]{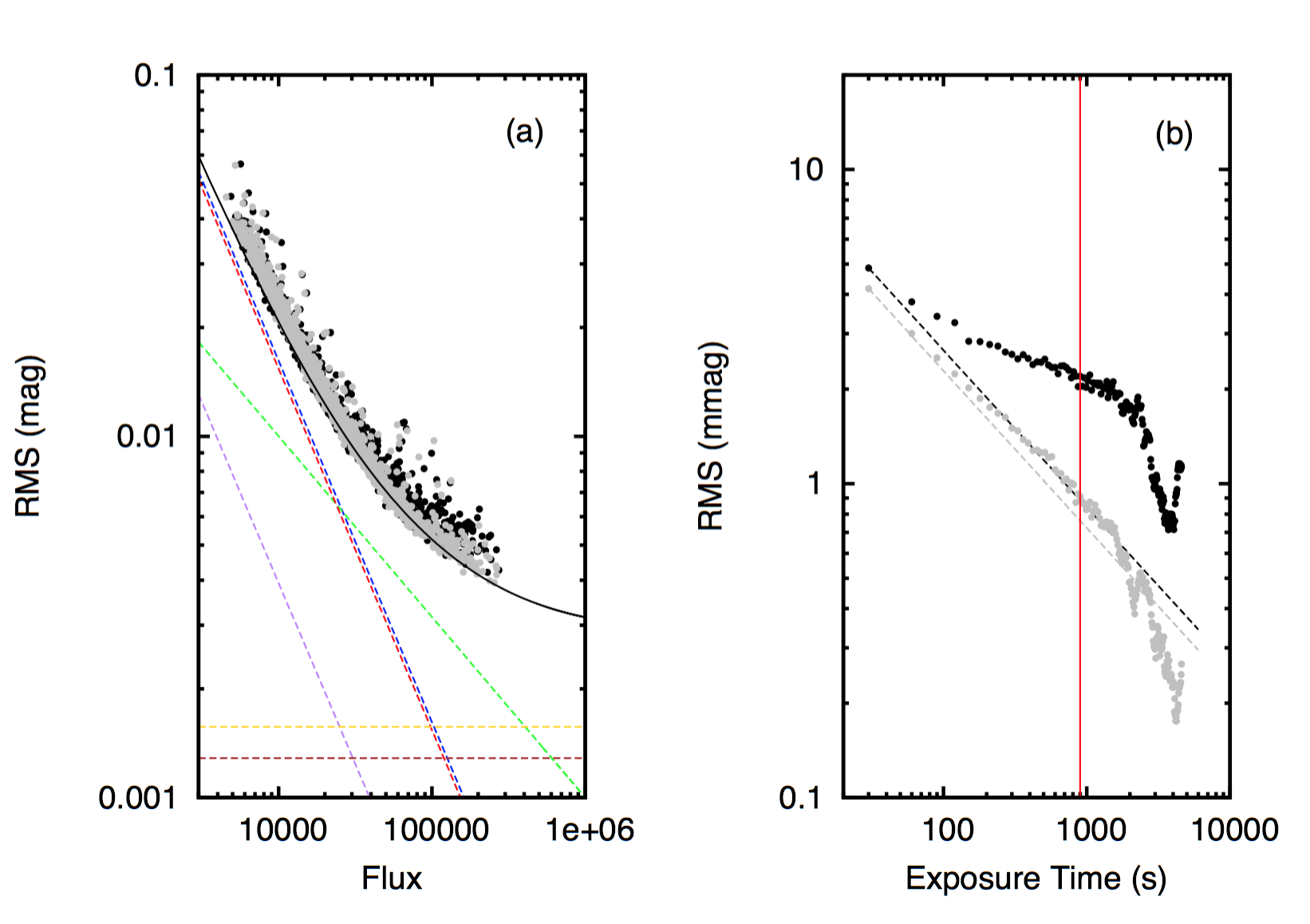}
\caption{Non linear effects of scattered light. Panel $a$ compares the RMS vs flux diagrams from 2009 November 19 with flat fielding (black points) and without (grey points). The green, blue, red, purple, yellow and brown dashed lines represent the noise models from the objects, sky, read noise, dark current, scintillation and flat fielding, respectively. The solid black line represents the total noise model. Panel $b$ shows the effects of flat fielding on the RMS vs binned exposure time plot of the $9$ brightest ($V~10.5$) non-varying stars. The black and grey dashed lines show the expected $1/\sqrt{n}$ decrease in RMS in the presence of Gaussian noise only, where $n$ is the number of points per bin. The solid red line highlights time where the binned RMS becomes sub-mmag ($\sim15$ min), which is well within a typical transit time scale.}
\label{figure:ffproblem}
\end{figure*}

We investigated the effects of systematic noise in the data by measuring the fractional improvement in the RMS of the noise in the light curves after the removal of up to $4$ trends using SysRem. Individual trends can be removed by overriding the stopping criterion of SysRem. We note that trends varying non-linearly with time (e.g. from scattered light) would not be identified by SysRem and hence remain in the light curves. The results after the forced removal of $1$, $2$, $3$ and $4$ trends can be seen in Fig \ref{figure:Tamuz1-4}. It is clear that the largest fractional improvement comes from correcting the atmospheric extinction caused by the airmass (see Fig \ref{figure:Tamuz1-4} panel b, red points) and that the removal of subsequent trends resulted in less significant improvements. Given that the stopping criterion for SysRem was typically satisfied after the removal of $1$ or $2$ trends only (see Fig \ref{figure:Tamuz1-4} red and black points, respectively) and that the RMS with binned exposure time follows a decrease as $1/\sqrt{n}$, we conclude that the system was relatively free from systematic trends. Panel (a) of Fig \ref{figure:ffproblem} also shows that the noise from NGTS can also be accounted for using a simple model made up from the typical sources of noise shown below:  

\begin{figure*}
\includegraphics[trim = 0mm 0mm 0mm 0mm, clip,width=7in]{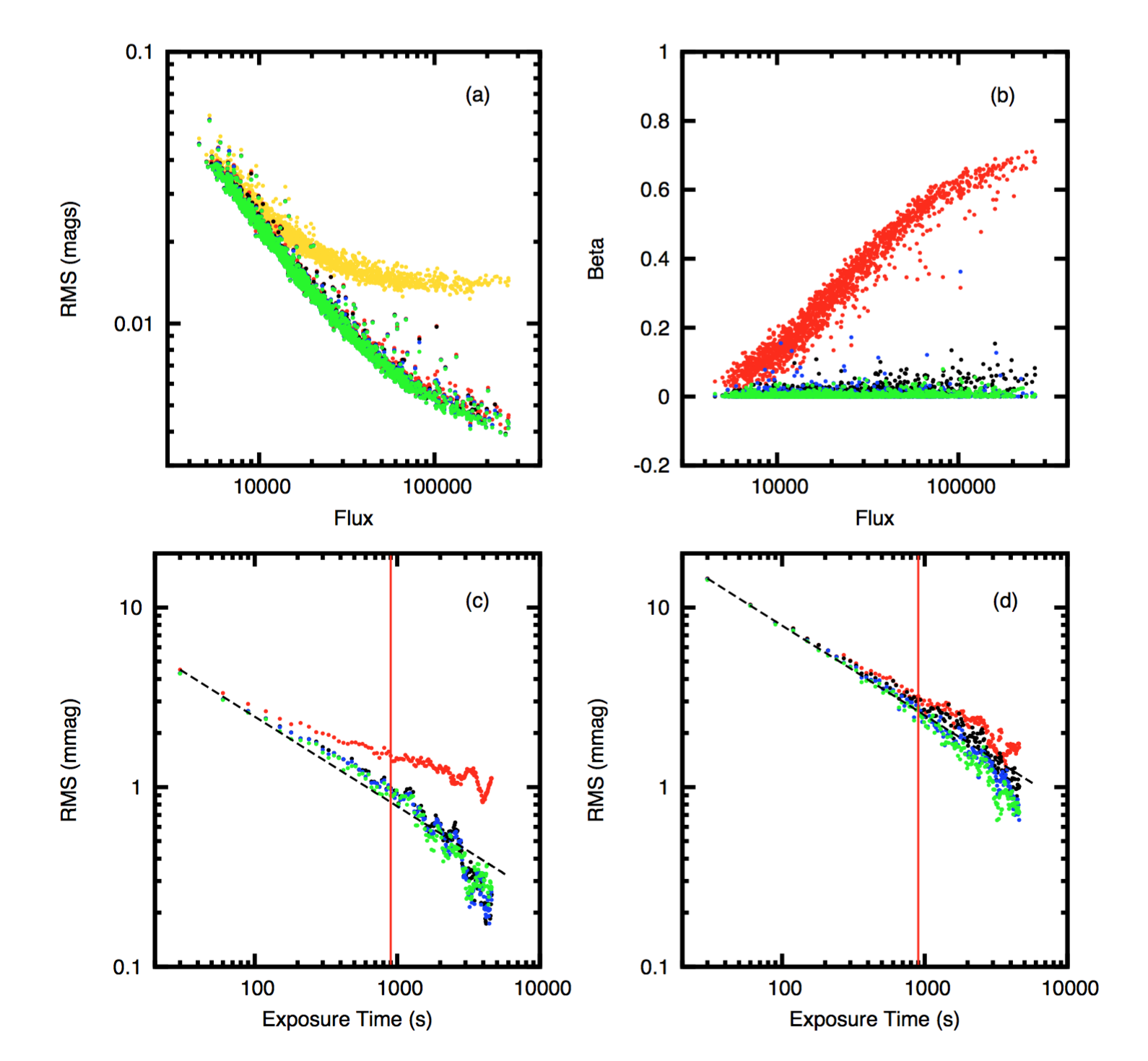}
\caption{Results from detrending NGTS-P data. The red, black, blue and green points represent the data after the removal of $1$, $2$, $3$ and $4$ trends with SysRem, respectively. The gold points represent the original data from 2009 September 19 uncorrected for airmass etc. Panel $a$ shows the different RMS vs flux diagrams for each level of detrending. Panel $b$ shows fractional improvement \emph{Beta} in each light curve's RMS versus its average flux for each level of detrending. Panels $c$ \& $d$ show the average RMS vs binned exposure time for the $9$ brightest stars ($V\sim10.5$) and $9$ stars of medium brightness ($V\sim12$) at each level of detrending, respectively. The black dashed lines show the expected $1/\sqrt{n}$ decrease in RMS in the presence of Gaussian noise only, where $n$ is the number of points per bin. The solid red line highlights the proposed $900$ s binned exposure time required to reach $1$ mmag RMS in the brightest sources.}
\label{figure:Tamuz1-4}
\end{figure*}

\begin{equation}
N_{\mathrm{total}} = \sqrt{f + f_{\mathrm{sky}} + DC + \left(N_{\mathrm{r}}\right)^{2} + N_{\mathrm{flat}} + N_{\mathrm{sc}}^{2}} \label{eq:TotalNoiseModel}
\end{equation}

\noindent where $f$, $f_{\mathrm{sky}}$, $DC$ and $N_{\mathrm{r}}$ are the flux from the target, sky, dark current and the read noise inside the photometry aperture, respectively and;

\begin{eqnarray}
N_{\mathrm{flat}}  &=& \frac{f \times n_{\mathrm{pix}}}{\sqrt{F_{\mathrm{total}}}} \\
N_{\mathrm{sc}} &=& 0.09D^{-2/3}\left(\sec\left(Z\right)\right)^{W}\mathrm{exp}\left(\frac{-h}{h_{\mathrm{0}}}\right)\left(2t\right)^{-1/2}
\end{eqnarray}

\noindent are the errors from flat fielding and the scintillation noise according to \citet{1998PASP..110..610D}, respectively. \mbox{$n_{\mathrm{pix}}$} is the number of pixels inside the photometric aperture, \mbox{$F_{\mathrm{total}}$} is the combined flux in \mbox{e$^{-}$ pix$^{-1}$} in the master flat field (typically \mbox{$\sim1\,000\,000$ e$^{-}$ pix$^{-1}$, if applied)}, $D$ is the diameter of the telescope aperture in cm, \mbox{$\sec\left(Z\right)$} is the airmass, $h$ is the altitude of the observatory in m, \mbox{$h_{\mathrm{0}} = 8\,000$ m} is the atmospheric scale height, $t$ is the integration time in s and $W$ is a variable dependent on angle between the line of sight and wind direction. \mbox{$W = 1.5$, $1.75$ or $2.0$} when observing perpendicular to the wind, close to the zenith and parallel to the wind, respectively. The agreement to a simple noise model strengthens further our conclusion that the system is relatively free from systematic noise.

\subsection{Known Transiting Planets}
\label{subsec:knownplanets}

\begin{deluxetable}{llccc}
\tabletypesize{\footnotesize}
\addtolength{\tabcolsep}{-2pt}
\tablecaption{Summary of GJ 1214b and GJ 436b observations with NGTS-P. The RMS quoted for each transit is that measured out of transit.}
\tablewidth{0pt}
\tablehead{
\colhead{Date} & \colhead{Object} & \colhead{Exposure} & \colhead{No.} & \colhead{Unbinned}\\
 & & \colhead{Time (s)} & \colhead{Images} & \colhead{RMS (mmag)} 
}
\startdata
2010 Apr 25	& GJ 1214b	& 55	& 219	& 6.65  \\
2010 May 14	& GJ 1214b	& 55	& 107	& 6.19  \\
2010 May 22	& GJ 1214b	& 55	& 215 	& 6.94  \\
2010 Jun 13	& GJ 1214b	& 60	& 199	& 6.38  \\
2010 Jun 21	& GJ 1214b	& 60	& 345	& 7.40  \\
2010 Jun 29	& GJ 1214b	& 60	& 152	& 5.93  \\
2010 Feb 04	& GJ 436b	& 10	& 579	& 3.66  \\
2010 Apr 27	& GJ 436b	& 12	& 800	& 9.72  \\
2010 May 05	& GJ 436b	& 18	& 399	& 3.12  \\
\enddata
\label{table:1} 
\end{deluxetable}

Observations of several known transiting planets with a range of transit depths and host star brightnesses were made with NGTS-P in order to characterise the system's primary function. We present the results from two example cases below, GJ 1214b \citep{2009Natur.462..891C} and GJ 436b \citep{2004ApJ...617..580B, 2007A&A...472L..13G}. NGTS-P also aided in the discovery of two new transiting exoplanets from the SuperWASP survey; the joint discovery of HAT-P-14/WASP-27b \citep{2010ApJ...715..458T,2011AJ....141..161S} and WASP-38b \citep{2011A&A...525A..54B}. 

To obtain initial system parameters for GJ 1214b and GJ 436b we have analysed the set of light curves for each planet using the EXO-NAILER code \citep{2016ApJ...830...43E}. EXO-NAILER\footnotemark\footnotetext{https://github.com/nespinoza/exonailer} is built on the BATMAN\footnotemark\footnotetext{https://github.com/lkreidberg/batman} exoplanet transit modelling code \citep{2015PASP..127.1161K}, and EMCEE\footnotemark\footnotetext{https://github.com/dfm/emcee}; an affine invariant Markov Chain Monte Carlo (MCMC) ensemble sampler \citep{emcee}. In an attempt to simulate a new discovery by NGTS - where the system parameters are initially unknown - we have used a wide range of uniform priors when fitting each planet. EXO-NAILER was run with a white noise only model, no light curve resampling and a square root limb darkening law from \citet{2013MNRAS.435.2152K}. Each MCMC run used $500$ walkers with $500$ jumps and $500$ steps during burn-in. For more details on EXO-NAILER the reader is referred to \citet{2016ApJ...830...43E}. Table \ref{table:exonailer} shows the priors and the results from fitting each planet. While the parameters in Table \ref{table:exonailer} agree with those published in the literature, we note that for Neptunes-sized or smaller exoplanets, higher precision follow-up light curves would be required to derive the system parameters more precisely.

\subsubsection{GJ 1214b}
\label{subsubsec:GJ1214b}

The first transiting super-Earth, GJ 1214b, was discovered by \citet{2009Natur.462..891C} as part of the MEarth Project \citep{2008PASP..120..317N,2009AIPC.1094..445I}, which, since 2007, has systematically targeted individual M type stars in the northern hemisphere with the goal of discovering small transiting exoplanets. At the time of observation with NGTS-P, GJ 1214b was amongst the smallest known transiting planets ($R_{p}=2.74^{+0.06}_{-0.05}$ R$_{\oplus}$; \citealt{2011ApJ...731..123K}), making it an interesting target for transit detection. However, GJ 1214 is a relatively faint M4.5V star at $V=14.71$ and therefore represents the faint extreme for a bright wide-field survey like NGTS.  

\begin{figure}
\includegraphics[scale=0.58,angle=0, trim=5mm 0mm 0mm 0mm]{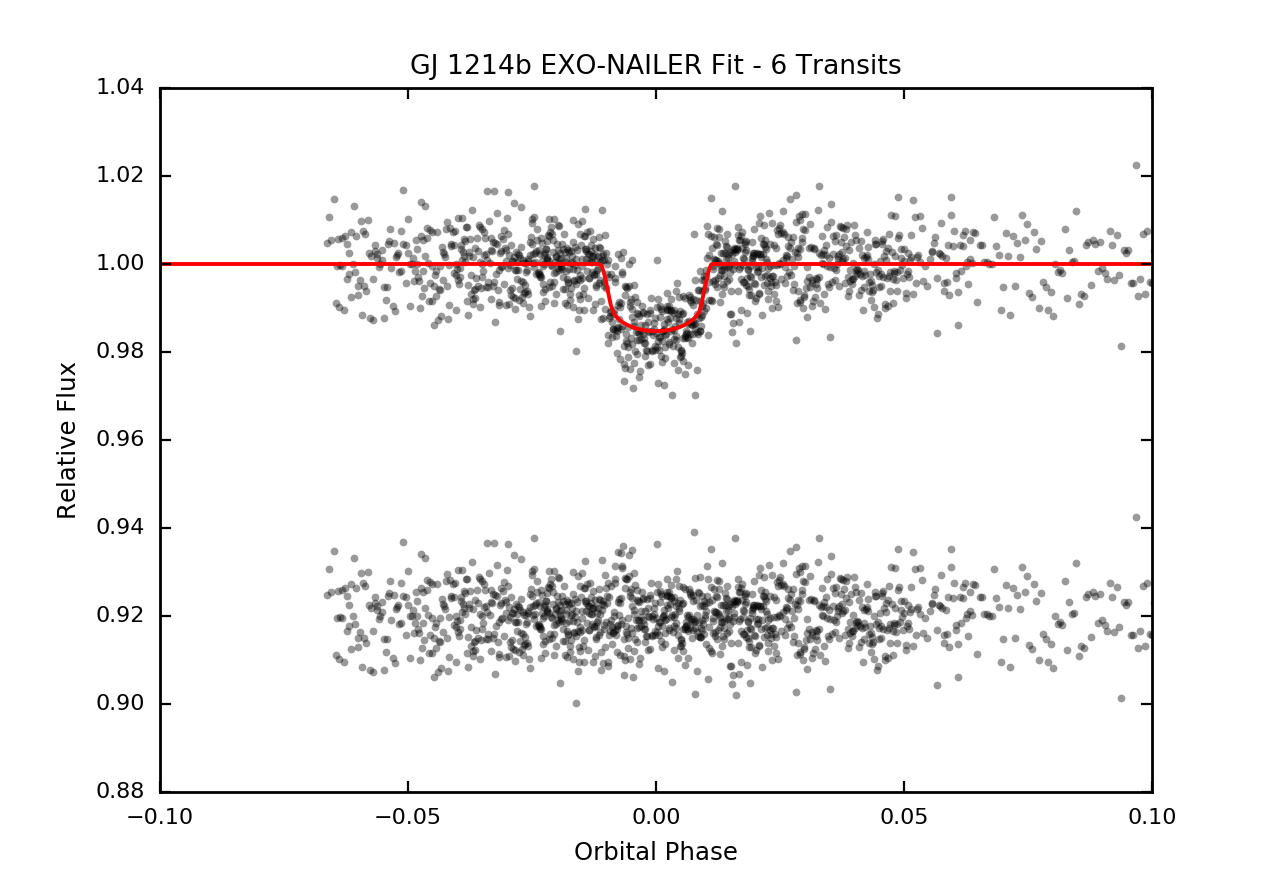}
\caption{An EXO-NAILER fit to 6 combined transits of GJ 1214b obtained with NGTS-P during the 2010 prototyping phase. The data are over plotted with the best fitting \citet{002ApJ...580L.171M} transiting planet model. The residuals after subtracting the best fitting model are shown below. The residuals have been offset by 0.08 for clarity.}
\label{figure:gj1214}
\end{figure}

Since the discovery of GJ 1214b the system has been the subject of intense scrutiny across many wavelengths and it has been the subject of several searches for additional planets via TTVs. \citet{2011ESS.....2.4009B} obtained $8$ transits of GJ 1214b in the $g$, $r$, $i$, $I$ and $z$ bands in the optical and $K_{s}$ ($\lambda=2.15\mu m$) and $K_{c}$ ($\lambda=2.27\mu m$) bands in the NIR in order to characterise any wavelength dependence on the planet-to-star radius ratio. Their observations show no evidence for wavelength dependence on the ratio of radii within the errors, apart from the $g$ band for which the ratio is slightly larger at the $\sim2\sigma$ level. \citet{2011ApJ...736...12B} presented a stellar variability analysis of several years of data, measuring  a peak-to-peak variability of $1$\% and a rotation period of $52.7\pm5$ d. Star spots have also been detected during transit \citep{0004-637X-730-2-82,2011ApJ...731..123K}. The presence of spots could also explain the difference in the planet-to-star radius ratio of \citet{2011ESS.....2.4009B} in the $g$ band, where the contrast in temperature between the stellar surface and the spots is more pronounced. Numerous observations of the transmission spectrum of GJ 1214b have been made spanning the optical and NIR regimes \citep{2010Natur.468..669B,2011ApJ...731L..40D,2011ApJ...736..132C,2011ApJ...736...78C,2012ApJ...747...35B}, all of which have shown to be essentially flat and featureless. 

Six transits of GJ 1214b were observed with NGTS-P in 2010. A summary of the observations and system parameters are given in Tables \ref{table:1} and \ref{table:exonailer}. The combined transits are shown in Fig \ref{figure:gj1214} over-plotted with the best fitting \citet{002ApJ...580L.171M} transiting planet model from EXO-NAILER. The transits of GJ 1214b were easily identifiable in NGTS-P data. We searched for Transit Timing Variations (TTVs) in the GJ 1214 system but found no significant deviation from the linear ephemeris of \citet{2009Natur.462..891C}, \citet{2011ApJ...731..123K} and \citet{2011ApJ...736...12B}.

\subsubsection{GJ 436b}
\label{subsubsec:GJ436b}
The warm-Neptune GJ 436b was originally discovered via the RV method \citep{2004ApJ...617..580B} and was subsequently observed to transit \citep{2007A&A...472L..13G}. Given that the circularisation time ($t_{\mathrm{circ}}\approx10^{8}$ yr) is much less than the age of the system ($6\times10^{9}$ yr), GJ 436b has a surprising non-zero eccentricity $e\sim0.14$ \citep{2007PASP..119...90M, 2007A&A...472L..13G,2007ApJ...667L.199D}. Several explanations have been proposed to explain the inflated eccentricity,  \citet{2012A&A...545A..88B} propose a 2-stage migrational process using perturbed Kozai cycles and a subsequent drastic decay of the orbital semi-major axis. Such a migrational process significantly increases $t_{\mathrm{circ}}$, meaning we are currently observing the second stage of migration where a significant eccentricity remains. 

\begin{figure}
\includegraphics[scale=0.58,angle=0, trim=5mm 0mm 0mm 0mm]{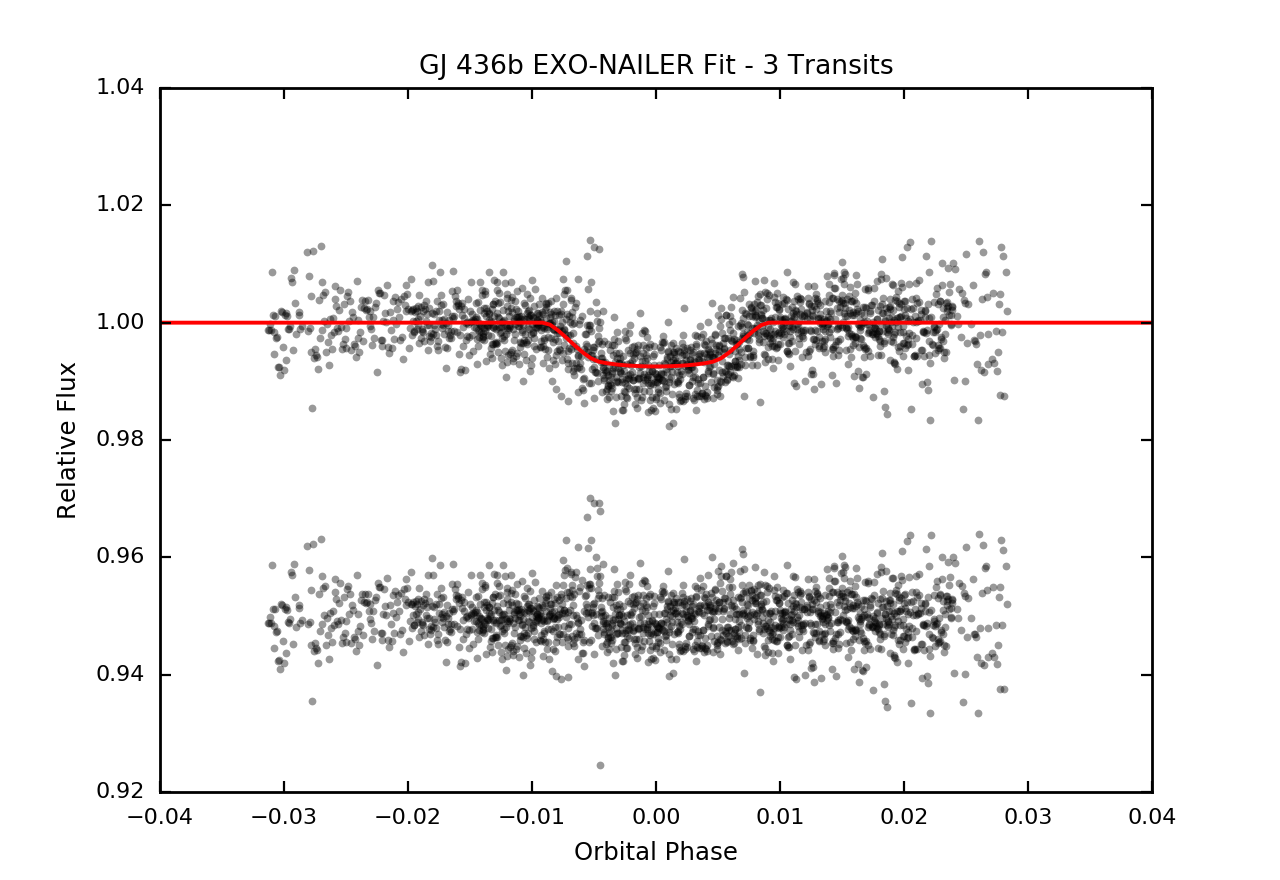}
\caption{An EXO-NAILER fit to 3 combined transits of GJ 436b obtained with NGTS-P during the 2010 prototyping phase. The data are over plotted with the best fitting \citet{002ApJ...580L.171M} transiting planet model. The residuals after subtracting the best fitting model are shown below. The residuals have been offset by 0.05 for clarity.}
\label{figure:gj436}
\end{figure}

With an M$2.5$\,V host star of brightness, planetary radius and transit depth $V=10.68$ \citep{2007ApJ...671L..65T}, $R_{p}=1.06\pm0.06$ $R_{\mathrm{Neptune}} $\citep{2010MNRAS.408.1689S} and $9$ mmag respectively, GJ 436b is a prime example of an NGTS-type target. Therefore, this object was observed during the prototyping phase to show the detection of such warm-Neptune sized exoplanets is possible with an NGTS-like system.  

Three transits of GJ 436b were observed with NGTS-P. A summary of the observations and system parameters are given in Tables \ref{table:1} and \ref{table:exonailer}. The combined transits are shown in Fig \ref{figure:gj436} over-plotted with the best fitting \citet{002ApJ...580L.171M} transiting planet model from EXO-NAILER. The transit obtained on 2010 April 27 was observed in poor weather conditions (high cloud and humidity). There was a $\sim10$ min gap during ingress when the high humidity forced us to close. This resulted in an increased RMS out of transit for that night (see Table \ref{table:1}). The transits of GJ 436b are clearly visible in our data and at a sufficient level of photometric accuracy to allow for initial characterisation of the planetary system.

\begin{deluxetable*}{lccc}
\tabletypesize{\scriptsize}
\addtolength{\tabcolsep}{-2pt}
\tablecaption{Summary of GJ 1214b and GJ 436b EXO-NAILER fitting priors and measured system parameters with comparison to published parameters. Priors on $P$ are set at the integer day boundaries of the known period, $T_{\mathrm{0}}$ ranges between $\pm1$ hour of mid-transit point on the first night of observation and $a/R_{\mathrm{star}}$, $R_{\mathrm{planet}}/R_{\mathrm{star}}$ and $i$ assume the wide ranges quoted below. $q1$ and $q2$ are the limb darkening coefficients for square-root law from \citet{2013MNRAS.435.2152K}. In both cases we fix $e=0$ and $\omega=90^{\circ}$. The published values for GJ 1214b and GJ 436b are assumed as the default parameters from the NASA Exoplanet Archive and are from \citet{2013A&A...549A..10H} and \citet{2014AcA....64..323M}, respectively. The value for $R_{\mathrm{planet}}/R_{\mathrm{star}}$ for GJ 1214b below is from \cite{2012ApJ...747...35B} as \citet{2013A&A...549A..10H} does not provide this parameter in the archive.}
\tablewidth{0pt}
\tablehead{
\colhead{Parameter} & \colhead{Prior} & \colhead{Posterior Value} & \colhead{Published Value}
}
\startdata
\bf{GJ 1214b:} & & \\
\hspace{1em}{$P$ (days)} & $\mathcal{U}$(1.0,2.0) & $1.58038\pm{0.00002}$ & $1.58040456\pm0.00000016$ \\
\hspace{1em}{$T_{\mathrm{0}}$-2450000 (BJD$_{\mathrm{TDB}}$)} & $\mathcal{U}$(5312.59384,5312.67384) & $5312.63375^{+0.00062}_{-0.00057}$ & $5320.535733\pm0.000021$ \\
\hspace{1em}{$a/R_{\mathrm{star}}$} & $\mathcal{U}$(10,20) & $15.34^{+0.68}_{-1.07}$ & $14.0^{+0.8}_{-0.7}$ \\
\hspace{1em}{$R_{\mathrm{planet}}/R_{\mathrm{star}}$} & $\mathcal{U}$(0.01,0.2) & $0.113\pm0.003$ & $0.1160\pm0.0005^{*}$  \\
\hspace{1em}{$i$ (deg)} & $\mathcal{U}$(80,90) & $86.16^{+1.28}_{-0.49}$ & $88.17\pm0.54$ \\
\hspace{1em}{$q_{1}$} & $\mathcal{U}$(0,1) & $0.590^{+0.267}_{-0.275}$ & - \\
\hspace{1em}{$q_{2}$} & $\mathcal{U}$(0,1) & $0.372^{+0.267}_{-0.237}$ & - \\
\bf{GJ 436b:} & & \\
\hspace{1em}{$P$ (days)} & $\mathcal{U}$(2.0,3.0) & $2.6439^{+0.0798}_{-0.1362}$ & $2.64388312\pm0.00000057$ \\
\hspace{1em}{$T_{\mathrm{0}}$-2450000 (BJD$_{\mathrm{TDB}}$)} & $\mathcal{U}$(5232.54273,5232.62273) & $5232.58423^{+0.00650}_{-0.00068}$ & $4510.80162\pm0.00007$ \\
\hspace{1em}{$a/R_{\mathrm{star}}$} & $\mathcal{U}$(10,20) & $12.43^{+1.56}_{-0.99}$ & $13.73^{+0.46}_{-0.43}$\\
\hspace{1em}{$R_{\mathrm{planet}}/R_{\mathrm{star}}$} & $\mathcal{U}$(0.01,0.2) & $0.089^{+0.011}_{-0.034}$ & $0.0822^{+0.0010}_{-0.0011}$\\
\hspace{1em}{$i$ (deg)} & $\mathcal{U}$(80,90) & $86.16^{+1.28}_{-0.49}$ & $86.44^{+0.17}_{-0.16}$ \\
\hspace{1em}{$q_{1}$} & $\mathcal{U}$(0,1) & $0.507^{+0.262}_{-0.226}$ & - \\
\hspace{1em}{$q_{2}$} & $\mathcal{U}$(0,1) & $0.362^{+0.309}_{-0.269}$ & - \\
\enddata
\label{table:exonailer} 
\end{deluxetable*}

\section{DISCUSSION} 
\label{sec:discussion}

The results presented in \S\ref{sec:performance} were very promising and subsequently led to the commencement of the full NGTS project. The prototyping phase uncovered several issues with our original design. These issues lead to a full review of the system design (CCD camera, detector, telescope and mount). The issues revealed during NGTS-P have subsequently been addressed for the full NGTS facility which has been recently commissioned at the ESO Paranal observatory, Chile. The issues and their solutions are discussed in more detail below.

A steady drift of $\sim4$ pixels h$^{-1}$ was seen in all of the NGTS-P data, even when autoguiding. After exhaustive investigation the source of this drift was never found. However, it was expected to come from a mechanical flexure between the science and autoguiding cameras. Hence we deemed an NGTS-like system with an off axis guider as unsuitable. To meet our science goal of sub-mmag photometry we require spatial stability over long periods of time at the sub-pixel level. To combat the sources of systematic noise we aim to fix stars to within $1$ pixel over time periods of weeks/months and possibly even years. To do so we created the versatile science-frame autoguiding algorithm DONUTS \citep{2013PASP..125..548M}. The algorithm has the added benefit of being able to guide on defocused PSFs so it can also be used at larger facilities to conduct even higher precision follow-up photometry after the initial discoveries by NGTS. The DONUTS algorithm is currently available as a Python package\footnotemark\footnotetext{https://github.com/jmccormac01/Donuts} \footnotemark\footnotetext{https://pypi.python.org/pypi/donuts}. A paper describing recent updates and improvements to the algorithm as well as extensive on-sky testing at NGTS on Paranal is currently being prepared (West et al., in prep).

At the beginning of the prototyping phase it became quickly apparent the telescope was suffering quite badly from scattered light during evening and morning twilight. A $30$ cm, light-weight foam baffle, covered in black flocking material, was installed on the telescope aperture to further minimise reflections. This reduced the intensity of the scattered light but its effects were still seen in our photometry. Figure \ref{figure:ffproblem} shows the effect of reducing a typical night's data with flat fielding (black points) and without (grey points). The initial difference between flat fielding and not, seen at the bright end of Fig \ref{figure:ffproblem}a, appears insignificant. However, as seen in Fig \ref{figure:ffproblem}b, measuring the RMS vs binned exposure time for the flat fielded data shows a significant systematic effect in the data and the points do not follow the line of $1/\sqrt{n}$ expected in the case of Gaussian noise only, where $n$ is the number of binned points. Excluding the flat fielding process while maintaining the remaining reduction steps removes the systematic trend. As the scattered light introduces noise in the data non-linearly, SysRem is unable to remove its effect, hence we chose not to flat field the data before analysis. Correct baffling of the final NGTS telescopes was therefore critical. 

\section{CONCLUSION} 
\label{sec:conclusion}

We have designed, built and tested a prototype system (NGTS-P) for the new, wide-field, transiting exoplanet survey NGTS. The goal of NGTS-P was to determine the level of systematic noise in an NGTS-like system and prove that millimagnitude photometry or better over a wide field could be achieved on transit time scales with an NGTS-like system. We have shown that millimagnitude photometry is possible using a modest aperture telescope and red-sensitive CCDs and that the noise in our system is essentially free from systematic effects. We have demonstrated our sensitivity to warm-Neptune and super-Earth sized exoplanets and also highlighted several key areas of NGTS which required more consideration. The prototyping phase has been a success and the lessons learned have been fed directly into the design of the full NGTS facility. NGTS is now in routine operation on Cerro Paranal, Chile, having achieved first light in January 2015. A detailed description of the final NGTS facility is currently being prepared (Wheatley et al. in prep). 

\section{ACKNOWLEDGMENTS} 
\label{sec:acknowledgments}

We would like to thank the anonymous referee for their constructive comments on this paper. \\

{\it Facilities: }\facility{NGTS-P}.

\end{document}